\documentclass[12pt]{article}

\usepackage{fancyheadings,a4,epsfig,pstricks,pst-node,pst-coil}
\usepackage{fancyheadings,a4,epsfig}
\usepackage[T1]{fontenc}
\usepackage{feynmf}
\renewcommand{\sectionmark}[1]%
 {\markright{\thesection\hspace{0.1cm} \ #1}}

\newcommand{\be}{\begin{equation}}
\newcommand{\ee}{\end{equation}}
\newcommand{\bea}{\begin{eqnarray}}
\newcommand{\eea}{\end{eqnarray}}

\newcommand{\ep}{i\varepsilon}
\newcommand{\nn}{\nonumber}

\newcommand{\dl}{\delta}

\begin{document}

\thispagestyle{empty}

\begin{center}{\bf {Natural Breaking of  Supersymmetry}}
\end{center}

\begin{center}{\bf{Vladim\'{i}r \v{S}auli}}\end{center}

\begin{center}{\it Dept. Theor. Phys. INP \v{R}e\v{z} near Prague, AVCR}
\end{center}

\begin{center}{Abstract}
\end{center}

\bigskip

The loop structure of two point Green's functions is investigated in  
the Wess-Zumino model in the formalism where the auxiliary fields are  
integrated out. In the usual frame of  perturbation theory 
the deviation from the non-renormalization theorem  is explicitly shown.
It is shown that Ward identity are not satisfied in this approach.
Further we  go beyond perturbation theory 
by solving a system of regularized Schwinger-Dyson equations (SDEs).
The mass splitting between fermions and bosons, which was already observed 
in perturbation theory level, is further enhanced.

\bigskip

\section{Introduction}

In this paper, we investigate the loop structure of the field theory model 
introduced by Wess and Zumino \cite{WESZUM1974} in formalism where the auxiliary
fields are integrated out. Our  aim is to explicitly check the ({\it in-})equivalence 
of quantum theories defined with and without the use of auxiliary fields.
The later approach (denoted as {\bf II} in the next text) is based on the quantization  of the Lagrangian model which 
has been obtained by eliminating auxiliary fields $F,G$ within the help of their  
classical equation of motion. The first is the well known manifest supersymmetry (Susy)
invariant (denoted as {\bf I} in the next text), which in its own has been examined
many times  for different reasons.

First let us ignore quantum correction and summarize the classical theory. 
In such case, the classical action is in our notation \cite{notation}
\bea
\label{action1}
S_C^I&=&\int d^4x {\cal{L_C^I}}\, ,  \hspace{10.2cm} \mbox{\bf I} 
\nn \\
{\cal{L_C^I}} &=&- {\scriptstyle {\frac{1}{2}}}(\partial A)^2 - 
{\scriptstyle {\frac{1}{2}}}(\partial B)^2 - {\scriptstyle {\frac{i}{2}}} 
\bar{\psi}\partial \!\!\!/ \psi + {\scriptstyle {\frac{1}{2}}} F^2 + 
{\scriptstyle {\frac{1}{2}}} G^2 \nonumber  \\
&+&  m(FA + GB - {\scriptstyle {\frac{1}{2}}}  \bar{\psi} \psi )  
\nonumber \\
&+& g(FA^2 - FB^2 + 2GAB -  \bar{\psi} \psi A - i \bar{\psi} 
\gamma_5 \psi B)\, ,
\label{clasI}
\eea
where classical Lagrangian ${\cal{L_C^I}}$ involves a scalar field $A$, 
a pseudoscalar field $B$, an anti-commuting Majorana spinor $\psi$, 
and two auxiliary scalar $F$ and pseudoscalar field $G$ both of canonical
dimension $mass^2$. Under supersymmetry transformation  each 
line in Lagrangian ${\cal{L_C^I}}$ transforms as a total  derivative, 
composing thus $S_C^I$ from three independent Susy invariant terms. 

Using a classical field equation of motion for $F,G$ one can obtain the 
following classical action  
\bea
S_C^{II}&=&\int d^4x {\cal L_{C}^{II}}   \hspace{10cm} \mbox{\bf II}
\nn \\
{\cal L_{C}^{II}}&=&\frac{-1}{2}\left[\partial_{\mu}A\partial^{\mu}A \label{klasik}
+\partial_{\mu}B\partial^{\mu}B
+i\bar{\psi}\not\partial\psi
+m^2A^2+m^2B^2+m\bar{\psi}\psi\right]
\nn \\
&-&mgA(A^2+B^2)-{\scriptstyle\frac{g^2}{2}}(A^2+B^2)^2
-g\bar{\psi}\psi A-ig\bar{\psi}\gamma_5\psi B\,
\label{clas2} .
\eea
It is widely believed that  theories {\bf I} and {\bf II}
should describe the same physics even when they are quantized, 
i.e. it is assumed that the equations of motion for the fields $F,G$
survive quantization.

In the paper \cite{ILIZUM1974} the classical theory {\bf I} was
quantized via path integral and considered generating functional for the Green's 
functions  is 
\be \label{generator}
Z^I[J^I]=\int {\cal{D}}A{\cal{D}}B{\cal{D}}F{\cal{D}}G{\cal{D}}\psi
\exp^{\frac{i}{\hbar}S_Q^I} \, ,
\ee
where the quantum action is defined as usually
\be 
S_Q^I=S_C^I+\int d^4x[A J_A+B J_B+F J_F+G J_G+\bar{\eta}\psi]\, ,
\ee
where  we have introduced space time dependent function $J^I(x)=(J_A,J_B,J_F,J_G,\eta)$ 
associated with the field $\Phi=(A,B,F,G,\psi)$, where $J^I$ includes c-number 
external sources of bosons and 
$\eta$ is an anticommuting Majorana spinor.
Imposing the transformation rules for $J^I$ then  the Susy can be retained 
and the well known  Ward identities can be write down
(for Susy Ward identities  see original paper \cite{ILIZUM1974}. 
 
Generating functional for the model {\bf II} is given by
\bea  \label{theory2}
Z^{II}[J^{II}]&=&\left<0\left|T\, exp \left\{\, i \,
\int d^4x \left[ J_A.A+J_B.B-\bar{\eta}\psi\right]\right\}\right|0\right>\, ,
\eea
where in contrary to the previous case , the Lagrangian of the model 
II has the ordinary  degrees of freedom, i.e the all fields have 
the usual kinetic terms. As consequence of elimination $F,G$
the all fields have now non-zero 
 canonical conjugated variable with non-vanishing  equal-time canonical 
(anti)commutation relations. From this, using a standard treatment 
 one can arrive to the alternative expression for the functional  (\ref{theory2})
\bea 
Z^{II}[J^{II}]&=&\frac{1}{Z[0]}\int {\cal{D}}A{\cal{D}}B{\cal{D}}\psi
\exp\left\{i \, S_Q^{II}\right\}
\nn \\
S_Q^{II}&=&S_C^{II}+\int d^4x \left[ J_A.A+J_B.B-\bar{\eta}\psi\right]\,
\eea
where $J_{II}=(J_A,J_B,\eta)$ are classical sources of Wess-Zumino  physical multiplet 
$\phi=(A,B,\psi)$.

Assuming no anomaly in {\bf I}, the Ward identities
\be
\Pi_A(p^2)=\Pi_B(p^2)=p^2\Gamma_1(p^2)
\ee
should be respected by the  renormalization.
In general these identities    constrain   various renormalization constants. Furthermore 
many of these constants are not actually necessary.
Already in the paper \cite{ILIZUM1974}  the non-renormalization theorem (NT)
for  WZM was proposed. 
This NT simply state that renormalized and bare quantities are related as 
\be \label{nonren}
m_r=Zm; \quad g_r=Z^{\frac{3}{2}}g
\ee
where $Z$ is the field strength  renormalization constant which is common to all field, 
i.e. $\psi=Z^{1/2}\psi_r$ and the same is valid for the fields $A$ and $B$; 
$m_r$ is renormalized mass and $g_r$ is the renormalized coupling constant.
 In the words the equations  (\ref{nonren}) means that all
diagrams belonging to vertices should be finite and the propagators could be 
divergent in very uniform way. In this place we should stress that the NT, ie. 
the actual finiteness of vertex functions, are proofed by an explicit evaluation 
of diagrams in superspace \cite{FULA1975,GRSIRO1979,GRSI1982}, (see also \cite{ILIZUM1974} for an indication  and for instance \cite{SEZGIN79} for the renormalization without supergraph technique)
 i.e. the appropriate proof is unavoidable based on the auxiliary field formalism represented by {\bf I}. 

Up to my knowledge a proper examination of infinities remains undertaken
in Susy field theories  without the explicit 
help of auxiliary field formulation. However, 
dealing with theory  with true physical degrees of freedom pertained, 
i.e. with the functional {\bf II}, we are not able to reproduce the NT at all. 
In the next section we will  demonstrate the deviation from the WTI and the  NT
explicitly  already at one loop.   

Although, we observe that one common $Z$ is still sufficient to renormalize all 
field content of WZM, the explicit calculation  suggests that the masses of boson and 
fermion sector necessarily differ due to the  quantum corrections. Explicitly, 
at one loop level we need to introduce the boson mass renormalization function $Z_{m_b}$,
 such that
\be
m^b=\frac{Z_{m_b}}{Z}\, m^b_r, ; \quad b=A,B
\ee
where $m^A_r=m^B_r$ are  renormalized boson masses when $Z_{m_A}=Z_{m_B}$ is naturally taken.
Remind here for completeness, 
that the relation $m_{\psi}=m_{\psi}/Z$ is still satisfied since the field $F,G$
do not couple to $\psi$. {\it In order to renormalize the  WZM {\bf II}, we must add the counterterms which break 
the original supersymmetry invariance.} Susy, if formulated by 
the quantum action in {\bf II}, does not survive quantization .

The paper is organized as follows. In the next section we write down the one 
loop perturbative result for selfenergies. In the section 3 we go beyond  the frame of
 perturbation theory 
and derive the set of Schwinger-Dyson equations (SDEs). After a simple truncation 
of SDEs we use some regularization techniques to solve  the SDEs numerically. 
Proposed method offers a simple estimate of Green's functions behavior  without 
loosing  connection with the 'bare' Lagrangian.

\section{1-loop corrections}
\label{oneloop}

For the discussion of  loop structure and the  renormalization  it is useful to define 
generating functional of one particle irreducible (1-PIR) Green's functions.
 This is obtained from the generating functional $W$ of connected Green function via 
Legendre transformation

\bea  \label{cornwall}
\Gamma[R]&=&
W[J]-\int d^4x\left(J_A.A+J_B.B-\bar{\eta}\psi\right)
\nn \\
W[J]&=&-ilogZ[J]\, ,
\eea
where $R$ are the semiclassical vacuum expectation of $\phi$
in the presence of $J$ already defined in the introduction.
They satisfy 
\bea \label{sources}
R_A&=&\frac{\delta W}{\delta J_A},R_B=\frac{\delta W}{\delta J_B}
,R_{\psi}=\frac{\delta W}{\delta \bar{\eta}},
\\
J_A&=&\frac{\delta W}{\delta R_A},J_B=\frac{\delta W}{\delta R_B}
,\eta=-\frac{\delta W}{\delta R_{\bar{\psi_c}}}.
\eea

First of all, let us mention that in the formulation {\bf I}  fermions
 does not interact with auxiliary field.
Therefore there is no difference when compare one loop fermion selfenergy in {\bf II} and 
in  {\bf I} (starting from (\ref{cornwall}) we will  omitted index II  since from 
now we will deal only with the theory {\bf II} and no confusion threats ).

 In addition we start  with the calculation of the propagator $G_B$. 
The calculation of $G_A$ will proceed similar steps. 
(We will relegate the appropriate  derivation of the propagator $G_A$ 
in fully nonperturbative fashion to  the next Section. 
The perturbation result then easily follow).
Before doing this explicitly we conventionally define the proper two point function 
\bea
\Gamma_B(x-y)&=&\left.\frac{\delta^2\Gamma[R]}{\delta R_B(x)\delta R_B(y)}\right|_{J=0}
=\int \frac{d^4p}{(2\pi)^4} e^{ip.(x-y)} \Gamma_B(p^2),
\nn \\
\Gamma_B(p^2)&=&\Gamma_B^{clas}-\Pi_B(p)=\Gamma_B^{clas}+\Gamma_B^{1loop}+...
\nn \\
\Gamma_B^{clas}&=&p^2-m^2,
\eea
(and the same we define  the propagator of $A$) noting that the identities 
$G_B^{-1}(p)=\Gamma_B(p)$,$G_A^{-1}(p)=\Gamma_A(p)$  are exactly valid even in the presence of external sources. The fermion propagator is defined as
\be \label{propag}
S^{-1}(p)=\not p\Gamma^{(1)}_{\psi}(p)+\Gamma_{\psi}^{(2)}(p)
=\not p-m-\Sigma(p)\, ,
\ee
where $\Gamma_1,\Gamma_2$ are two independent function and $\Sigma(p)$ is the conventional selfenergy. 
Equivalently, it is sometimes convenient to introduce the following formula
\be
S(p)=\frac{{\cal{F}}(p)}{\not p -{\cal{M}}(p)} \quad ,
\ee
where the so called renormalization wave function is defined through the eq.: 
${\cal{F}}^{-1}=\Gamma^{(1)}$ and 
 mass function is ${\cal{M}}=\Gamma^{(2)}/\Gamma^{(1)}$.

The appropriate contribution to  $\Pi_B$ can be represented as
\begin{fmffile}{fskalar2}
\bea \label{bloop}
&&\Pi_B(p^2)=
\parbox{2.0\unitlength}{%
\begin{fmfgraph}(2.0,2.0)
\fmfpen{thick}
\fmfleft{i}
\fmfright{o}
\fmf{dots}{v,v}
\fmf{dots}{i,v,v,o}
\end{fmfgraph}}
\, \, + \, \,  
\parbox{2.0\unitlength}{%
\begin{fmfgraph}(2.0,2.0)
\fmfpen{thick}
\fmfleft{i}
\fmfright{o}
\fmf{dashes}{v,v}
\fmf{dots}{i,v}
\fmf{dots}{v,o}
\end{fmfgraph}}
\, \, + \, \,  
\parbox{2.0\unitlength}{%
\begin{fmfgraph}(2.0,2.0)
\fmfpen{thick}
\fmfleft{i}
\fmfright{o}
\fmf{dots}{i,v1}
\fmf{dots}{v2,o}
\fmf{dots,left,tension=0.25}{v1,v2}
\fmf{dashes,right,tension=0.25}{v1,v2}
\end{fmfgraph}}
\, \, + \, \,  
\parbox{2.0\unitlength}{%
\begin{fmfgraph}(2.0,2.0)
\fmfpen{thick}
\fmfleft{i}
\fmfright{o}
\fmf{dots}{i,v1}
\fmf{dots}{v2,o}
\fmf{fermion,left,tension=0.25}{v1,v2}
\fmf{fermion,left,tension=0.25}{v2,v1}
\end{fmfgraph}}
\nn \\
&&=
-i\int\frac{d^4l}{(2\pi)^4}
\left\{\left[\frac{6g^2}{l^2-m^2+\ep}\right]+\left[\frac{2g^2}{l^2-m^2+\ep}\right]\right.
\nn \\
&&+\left[\frac{4m^2g^2}{((l-p)^2-m^2+\ep)(l^2-m^2+\ep)}\right]
\nn \\
&&-\left.\left[\frac{8g^2}{l^2-m^2+\ep}
+\frac{-4g^2p^2}{((l-p)^2-m^2+\ep)(l^2-m^2+\ep)}\right]\right\},
\eea
\end{fmffile}
where each term in the each bracket $[]$ in (\ref{bloop}) is associated with 
given Feynman diagram depicted in the first line (in given order). 
The dashed (dot) line  corresponds with the propagator of A(B) field,
 the solid line stands for the fermion propagator.
Convenient shift and some obvious algebra has been done 
( we clear out the numerator $l.p$ by the use of the identity 
$2l.p=G^{-1}(l)-G^{-1}(l-p)+p^2$) in expression for the fermion loop. 
 The appropriate cancellation of quadratic divergences between scalar tadpoles 
and fermion loop is obvious.
Summing all together, the complete one loop result is given by the following 
logarithmically divergent expression  
\bea \label{simple}
\Pi_B(p)&=&\frac{\alpha}{\pi}(p^2+m^2){\cal{I}}(p)
 \\
I(p^2)&=&-i\int\frac{d^4l}{\pi^2}\frac{1}{((l-p)^2-m^2+\ep)(l^2-m^2+\ep)}\, 
\nn \\
&=&\lim_{\Lambda\rightarrow\infty}\int_{4m^2}^{\Lambda^2} d\omega
\frac{\sqrt{1-\frac{4m^2}{\omega}}}{p^2-\omega+i\epsilon}\, ,
\label{berny}
\eea 
where we have defined $\alpha=g^2/(4\pi)$.

For completeness we also review the appropriate set of relevant Feynman diagrams which 
complete 
one loop contribution to $\Pi_A$. These are

\begin{fmffile}{fskalar3}
\bea \label{aloop}
&&\Pi_A(p^2)=
\parbox{2.0\unitlength}{%
\begin{fmfgraph}(2.0,2.0)
\fmfpen{thick}
\fmfleft{i}
\fmfright{o}
\fmf{dashes}{v,v}
\fmf{dashes}{i,v,v,o}
\end{fmfgraph}}
\, \, + \, \,  
\parbox{2.0\unitlength}{%
\begin{fmfgraph}(2.0,2.0)
\fmfpen{thick}
\fmfleft{i}
\fmfright{o}
\fmf{dots}{v,v}
\fmf{dashes}{i,v}
\fmf{dashes}{v,o}
\end{fmfgraph}}
\, \, + \, \,  
\parbox{2.0\unitlength}{%
\begin{fmfgraph}(2.0,2.0)
\fmfpen{thick}
\fmfleft{i}
\fmfright{o}
\fmf{dashes}{i,v1}
\fmf{dashes}{v2,o}
\fmf{fermion,left,tension=0.25}{v1,v2}
\fmf{fermion,left,tension=0.25}{v2,v1}
\end{fmfgraph}}
\nn \\
\, \, &+& \, \,  
\parbox{2.0\unitlength}{%
\begin{fmfgraph}(2.0,2.0)
\fmfpen{thick}
\fmfleft{i}
\fmfright{o}
\fmf{dashes}{i,v1}
\fmf{dashes}{v2,o}
\fmf{dashes,left,tension=0.25}{v1,v2}
\fmf{dashes,right,tension=0.25}{v1,v2}
\end{fmfgraph}}
\, \, + \, \,  
\parbox{2.0\unitlength}{%
\begin{fmfgraph}(2.0,2.0)
\fmfpen{thick}
\fmfleft{i}
\fmfright{o}
\fmf{dashes}{i,v1}
\fmf{dashes}{v2,o}
\fmf{dots,left,tension=0.25}{v1,v2}
\fmf{dots,right,tension=0.25}{v1,v2}
\end{fmfgraph}}
\eea
\end{fmffile}
The  evaluation of $\Pi_A$ simply gives 
\be
\Pi_A(p^2)=\Pi_B(p^2)\, .
\ee
And recall that in our notation the fermion selfenergy can be evaluated as
\be
\Sigma(p)=\not p\, \frac{\alpha}{\pi}\, I(p^2).
\ee

In  the limit $\Lambda\rightarrow \infty$ in (\ref{berny})  the infinite term  
$m^2\, \frac{\alpha}{\pi}I(p^2)$ in the expression (\ref{bloop}) for boson selfenergy 
can be removed only by renormalization of  the masses $m_A$ and $m_B$ in the WZM. 
The NT as well as Susy WI are not valid for theory {\bf II}. Assuming $g$ is a rather
 small number, i.e. $g<<1$ then the   perturbation theory inspection provides the
 strong evidence (if not proof) of  non-equivalent formulation of  the supersymmetric
 quantum field theory defined throughout {\bf I} and {\bf II}. Note, it does not mean that 
one of the the theories is incorrect. The result should be read as that the quantum field
 theory {\bf II} is not equivalent to the theory {\bf I} wherein the classical equation 
of motion for phantom fields 
$F,G$ are simultaneously satisfied (as it was supposed in \cite{WESZUM1974}).


\section{Masses beyond perturbation theory}

In this Section we derive Schwinger-Dyson equations (SDEs) for the propagators 
 of the WZM {\bf II}. These are the first of infinite system of
 quantum equations of motion which relate Green's functions of theory with themselves. 
If solved exactly, we would obtain the full 
information about the theory. The technical impossibility of such a procedure is obvious 
and we truncate the system of SDEs in a way that if we approximate the exact Green's 
functions inside the loops by the bare ones then the one-loop results obtained 
in the previous Section are reproduced.

Before presenting the derivation we anticipate the resulting equations graphically. 
The set of SDEs for two point Green's functions can be diagrammatically represented as:

\begin{fmffile}{fskalar}
\bea  \label{USDE}
&&G_A^{-1}(p^2)=p^2-m^2-\Pi_A(p^2) \, ,
\nn \\
&&\Pi_A(p^2)=
\parbox{2.0\unitlength}{%
\begin{fmfgraph}(2.0,2.0)
\fmfpen{thick}
\fmfleft{i}
\fmfright{o}
\fmf{dashes}{v,v}
\fmf{dashes}{i,v,v,o}
\end{fmfgraph}}
\, \, + \, \,  
\parbox{2.0\unitlength}{%
\begin{fmfgraph}(2.0,2.0)
\fmfpen{thick}
\fmfleft{i}
\fmfright{o}
\fmf{dots}{v,v}
\fmf{dashes}{i,v}
\fmf{dashes}{v,o}
\end{fmfgraph}}
\, \, + \, \,  
\parbox{2.0\unitlength}{%
\begin{fmfgraph}(2.0,2.0)
\fmfpen{thick}
\fmfleft{i}
\fmfright{o}
\fmf{dashes}{i,v1}
\fmfblob{0.1w}{v1}
\fmf{dashes}{v2,o}
\fmf{fermion,left,tension=0.25}{v1,v2}
\fmf{fermion,left,tension=0.25}{v2,v1}
\end{fmfgraph}}
\nn \\
&+&\, \,  
\parbox{2.0\unitlength}{%
\begin{fmfgraph}(2.0,2.0)
\fmfpen{thick}
\fmfleft{i}
\fmfright{o}
\fmf{dashes}{i,v1}
\fmfblob{0.1w}{v1}
\fmf{dashes}{v2,o}
\fmf{dashes,left,tension=0.25}{v1,v2}
\fmf{dashes,right,tension=0.25}{v1,v2}
\end{fmfgraph}}
\, \, + \, \,  
\parbox{2.0\unitlength}{%
\begin{fmfgraph}(2.0,2.0)
\fmfpen{thick}
\fmfleft{i}
\fmfright{o}
\fmf{dashes}{i,v1}
\fmfblob{0.1w}{v1}
\fmf{dashes}{v2,o}
\fmf{dots,left,tension=0.25}{v1,v2}
\fmf{dots,right,tension=0.25}{v1,v2}
\end{fmfgraph}}
\, \, +\, \, \parbox{2.5\unitlength}{%
\begin{fmfgraph}(2.5,2.0)
\fmfpen{thick}
\fmfleft{i}
\fmfright{o}
\fmf{dashes}{i,v1}
\fmfv{d.sh=diamond,d.f=hatched,d.si=0.15w,l=$\Gamma_4$}{v1}
\fmf{dashes}{v2,o}
\fmf{dashes,left,tension=0.2}{v1,v2}
\fmf{dashes,tension=0.2}{v1,v2}
\fmf{dashes,right,tension=0.2}{v1,v2}
\end{fmfgraph}}
\, \,  +  \, \,
\parbox{2.5\unitlength}{%
\begin{fmfgraph}(2.5,2.0)
\fmfpen{thick}
\fmfleft{i}
\fmfright{o}
\fmf{dashes}{i,v1}
\fmfv{d.sh=diamond,d.f=hatched,d.si=0.15w,l=$\Gamma_4$}{v1}
\fmf{dashes}{v2,o}
\fmf{dots,left,tension=0.2}{v1,v2}
\fmf{dots,tension=0.2}{v1,v2}
\fmf{dashes,right,tension=0.2}{v1,v2}
\end{fmfgraph}}
\nn \\
&+&\, \,  
\parbox{2.5\unitlength}{%
\begin{fmfgraph}(2.5,2.0)
\fmfpen{thick}
\fmfleft{i}
\fmfright{o}
\fmf{dashes}{i,v1}
\fmfblob{0.1w}{v2}
\fmfblob{0.1w}{v3}
\fmf{dashes}{v3,o}
\fmf{dashes,left,tension=0.2}{v1,v3}
\fmf{dots,left,tension=0.1}{v1,v2}
\fmf{dots,right,tension=0.1}{v1,v2}
\fmf{dashes,right,tension=0.2}{v2,v3}
\end{fmfgraph}}
\, \,  +  \, \,  
\parbox{2.5\unitlength}{%
\begin{fmfgraph}(2.5,2.0)
\fmfpen{thick}
\fmfleft{i}
\fmfright{o}
\fmf{dashes}{i,v1}
\fmfblob{0.1w}{v2}
\fmfblob{0.1w}{v3}
\fmf{dashes}{v3,o}
\fmf{dashes,left,tension=0.2}{v1,v3}
\fmf{dashes,left,tension=0.1}{v1,v2}
\fmf{dashes,right,tension=0.1}{v1,v2}
\fmf{dashes,right,tension=0.2}{v2,v3}
\end{fmfgraph}}
\label{selfa}
\eea
%
%
\bea  \label{USDE2}
&&G_B^{-1}(p^2)=p^2-m^2-\Pi_B(p^2) \, ,
\nn \\
&&\Pi_B(p^2)=
\parbox{2.0\unitlength}{%
\begin{fmfgraph}(2.0,2.0)
\fmfpen{thick}
\fmfleft{i}
\fmfright{o}
\fmf{dots}{v,v}
\fmf{dots}{i,v,v,o}
\end{fmfgraph}}
\, \, + \, \,  
\parbox{2.0\unitlength}{%
\begin{fmfgraph}(2.0,2.0)
\fmfpen{thick}
\fmfleft{i}
\fmfright{o}
\fmf{dashes}{v,v}
\fmf{dots}{i,v}
\fmf{dots}{v,o}
\end{fmfgraph}}
\, \, + \, \,  
\parbox{2.0\unitlength}{%
\begin{fmfgraph}(2.0,2.0)
\fmfpen{thick}
\fmfleft{i}
\fmfright{o}
\fmf{dots}{i,v1}
\fmfblob{0.1w}{v1}
\fmf{dots}{v2,o}
\fmf{dots,left,tension=0.25}{v1,v2}
\fmf{dashes,right,tension=0.25}{v1,v2}
\end{fmfgraph}}
\, \, + \, \,  
\parbox{2.0\unitlength}{%
\begin{fmfgraph}(2.0,2.0)
\fmfpen{thick}
\fmfleft{i}
\fmfright{o}
\fmf{dots}{i,v1}
\fmfblob{0.1w}{v1}
\fmf{dots}{v2,o}
\fmf{fermion,left,tension=0.25}{v1,v2}
\fmf{fermion,left,tension=0.25}{v2,v1}
\end{fmfgraph}}
\nn \\
&&\, \, +\, \,\parbox{2.5\unitlength}{%
\begin{fmfgraph}(2.5,2.0)
\fmfpen{thick}
\fmfleft{i}
\fmfright{o}
\fmf{dots}{i,v1}
\fmfv{d.sh=diamond,d.f=hatched,d.si=0.15w,l=$\Gamma_4$}{v1}
\fmf{dots}{v2,o}
\fmf{dots,left,tension=0.2}{v1,v2}
\fmf{dots,tension=0.2}{v1,v2}
\fmf{dots,right,tension=0.2}{v1,v2}
\end{fmfgraph}}
\, \,  +  \, \,
\parbox{2.5\unitlength}{%
\begin{fmfgraph}(2.5,2.0)
\fmfpen{thick}
\fmfleft{i}
\fmfright{o}
\fmf{dots}{i,v1}
\fmfv{d.sh=diamond,d.f=hatched,d.si=0.15w,l=$\Gamma_4$}{v1}
\fmf{dots}{v2,o}
\fmf{dots,left,tension=0.2}{v1,v2}
\fmf{dashes,tension=0.2}{v1,v2}
\fmf{dashes,right,tension=0.2}{v1,v2}
\end{fmfgraph}}
\, \,  +  \, \,  
\parbox{2.5\unitlength}{%
\begin{fmfgraph}(2.5,2.0)
\fmfpen{thick}
\fmfleft{i}
\fmfright{o}
\fmf{dots}{i,v1}
\fmfblob{0.1w}{v2}
\fmfblob{0.1w}{v3}
\fmf{dots}{v3,o}
\fmf{dashes,left,tension=0.2}{v1,v3}
\fmf{dots,left,tension=0.1}{v1,v2}
\fmf{dashes,right,tension=0.1}{v1,v2}
\fmf{dots,right,tension=0.2}{v2,v3}
\end{fmfgraph}}
\label{selfb}
\eea
\bea  \label{USDE3}
&&S^{-1}(p)=\not p-m-\Sigma(p) \, ,
\nn \\
&&\Sigma(p)=
\parbox{2.0\unitlength}{%
\begin{fmfgraph}(2.0,2.0)
\fmfpen{thick}
\fmfleft{i}
\fmfright{o}
\fmf{fermion}{i,v1}
\fmfblob{0.1w}{v1}
\fmf{fermion}{v2,o}
\fmf{dots,left,tension=0.25}{v1,v2}
\fmf{fermion,right,tension=0.25}{v1,v2}
\end{fmfgraph}}
\, \, \, +  \, \, \,
\parbox{2.0\unitlength}{%
\begin{fmfgraph}(2.0,2.0)
\fmfpen{thick}
\fmfleft{i}
\fmfright{o}
\fmf{fermion}{i,v1}
\fmfblob{0.1w}{v1}
\fmf{fermion}{v2,o}
\fmf{dashes,left,tension=0.25}{v1,v2}
\fmf{fermion,right,tension=0.25}{v1,v2}
\end{fmfgraph}}
\label{selfpsik}\, ,
\eea
\end{fmffile}
where now (in contrary to the previous Section) the all internal lines
inside the loops represent the fully dressed propagators. 
The diamonds and bloops stand for dressed quartic and trilinear vertices 
which satisfy their own SDE. The above mentioned truncation of SDEs 
simply means that in the equations (\ref{selfa}),(\ref{selfb}),(\ref{selfpsik})
 the two-loop diagrams should be  neglected and  dressed vertices should be 
 replaced by the classical ones.

The derivation of SDEs is tedious but necessary task we now turn.
For this purpose  we accommodate the elegant and waterproof functional method 
originally developed by Symanzik and others {\cite{SYMANZIK} }. 
Review  of this formalism  can be found in  some standard introductory textbook 
 (e.g. \cite{RIVERS}). The SDEs can be derived by further  variations of the 
following compact formula:
\bea  \label{qoe}
\frac{\delta{\Gamma[R]}}{\delta R_{\phi_i}(x)}=
\left.{\delta S_c}\over{\delta \phi_i(x)}\right|_{\phi(x)=\check{\phi}(x)}
\eea
where $\check{\phi}$ is  operator acting on the right
\be
\check{\phi_i}(x)=R_{\phi_i}(x)
-i\int d^4y \frac{\dl^2 W[J]}{\dl J_{\phi_i}(x)\dl J_{\phi_i}(y)}
\frac{\dl }{\dl R_{\phi_i}(y)}\, ,
\ee
where  $J_{\phi_i}$ is an associated external source
of $\phi_i$, $R_{\phi_i}$ is the classical expectation value of $\phi_i$. 
In words, the right handed differential operator
$\check{\phi_i}$  replaces the Heisenberg field operators $\phi$ positioned on the left
in the classical equation of motion. In order to obtain the physical Green's functions 
the sources are switched off at the end of the calculation.

Now we are going to explain some details 
for  the equation for $G_A^{-1}$. For the  case of space-like derivatives of $\phi=A$ 
we can use per-partes integration as usually
\bea  \label{forA}
\frac{\delta\Gamma[R]}{\delta R_{A}(x)}&=&
-\left\{ (\partial_{\mu} \partial^{\mu}+m^2)R_{A}(x)+g\bar{\psi}(x)\psi(x)\right.
 \\
&+&\left.mg(3\check{A}^2(x)+\check{B}^2(x))+
2g^2\check{A}(x)(\check{A}^2(x)+\check{B}^2(x))\right\}\, ,
\nn 
\\
\check{A}(x)&=&R_A(x)-
i\int d^4y \frac{\dl^2 W[J]}{\dl J_A(x)\dl J_A(y)}
\frac{\dl }{\dl R_A(y)}\, ,
\nn \\
\check{B}(x)&=&R_B(x)-
i\int d^4y \frac{\dl^2 W[J]}{\dl J_B(x)\dl J_B(y)}
\frac{\dl }{\dl R_B(y)}\, ,
\nn \\
\check{\psi}(x)&=&R_{\psi}(x)
-i\int d^4y \frac{\dl^2 W[J]}{\dl \bar{\eta}_{\psi}(x)\dl \eta_{\psi}(y)}
\frac{\dl }{\dl \bar{R}_{\psi}(y)}\, ,
\nn
\eea   
where we have used obvious identity
\be
\check{\phi}(x)1=R_{\phi}(x).
\ee 

In addition we will use convenient shorthand notation for the measures and 
dirac delta in Minkowski spacetime.
\be
\tilde{dx}=d^4x;\quad \tilde{dp}=\frac{d^4p}{(2\pi)^4}; \quad \delta_{xy}=\delta^4(x-y).
\ee
Also we introduce the abbreviation for the variations of the 
generating functionals:
\bea
\frac{\dl^n \Gamma[R]}{\dl R_{i_1}(x_1)...\dl R_{i_n}(x_n)}
&=&\Gamma_{(x_1...x_n)}^{R_{i_1},..., R_{i_n}};
\nn \\
\frac{\dl^n W[J]}{\dl J_{\phi_{i_1}}(x_1)...\dl J_{\phi_{i_n}}(x_n)}
&=&W_{(x_1...x_n)}^{\phi_{i_1},...,\phi_{i_n}}.
\eea

Once more variation of the Eq. (\ref{forA}) with respect to $R_A$
 gives 

\bea   \label{xadse}
&&\Gamma^{R_AR_A}_{(y,x)}=
-\partial_{\mu} \partial^{\mu}\delta_{xy}-m^2\delta_{xy}-\Pi_A(x,y)
 \\
&&\Pi_A(x,y)=\frac{\delta }{\dl R_A(y)}
\left\{mg[3\check{A}^2(x)+\check{B}^2(x)]\right.
\nn \\
&&\left.+2g^2\check{A}(x)[\check{A}^2(x)+\check{B}^2(x)]
+g\check{\bar{\psi}}(x)\check{\psi}(x)\right\}
\label{xadse2}
\eea

Switching the external sources $J$ the equation (\ref{xadse}) represents
the Fourier transformation of the inverse propagator $G_A^{-1}$
\be \label{hehe}
G_A^{-1}(x,y)=\Gamma^{R_AR_A}_{(y,x)}|_{R_{\phi}=0}=
\int \tilde{dp} e^{ip.(x-y)}\left[p^2-m^2-\Pi_A(p)\right]\, ,
\ee
since in our notation 
\be \label{huhu}
W^{AA}_{(y,x)}|_{J_{\phi}=0}=
-\int \tilde{dp} G_A(p) e^{ip.(x-y)}\, .
\ee

 Recall at this place  general functional identity 
\be \label{iden}
\int \tilde{dz} \Gamma^{R_AR_A}_{(y,z)}W^{AA}_{(z,x)} =-\delta_{xy}
\ee
which  is valid  in the case of the presence of the external sources $J$ to.
  
The cubic $A^3$ term in the Lagrangian  is responsible 
for the following contribution to $\Pi_A$ (i.e. the first term in (\ref{xadse2})): 

\bea \label{jednicka}
&&\frac{\dl \check{A}^2(x)}{\dl R_A(y)}1=\frac{\dl }{\dl R_A(y)}
\left[R_A^2(x)-i W^{AA}_{(x,x)}\right]
\nn \\
&&=2R_A(x)\dl_{xy}
-i\int \tilde{dz}\tilde{du} W^{AA}_{(x,z)} \Gamma^{R_AR_AR_A}_{(y,z,u)}W^{AA}_{(u,x)}
\eea

The second required variation in (\ref{xadse}) is
\bea    \label{dvojka}
&&\frac{\dl \check{B}^2(x)}{\dl R_A(y)}1=
\frac{\dl }{\dl R_A(y)}
\left[R_B^2(x)-i W_{(x,x)}^{BB}\right]
\nn \\
&&=-i \int \tilde{dz}\tilde{du} W^{BB}_{(x,z)} \Gamma^{R_AR_BR_B}_{(y,z,u)}W^{BB}_{(u,x)},
\eea
where we have repeatedly used
the relation
%
\be  \label{srpen}
\frac{\dl W_{(x_1,x_2)}^{\phi_1\phi_2}}{\dl \Omega(y)}=
\int \tilde{dz}\tilde{du} W^{\phi_1\phi_2}_{(x_1,z)} \Gamma^{\Omega R_{\phi_1}R_{\phi_1}}_{(y,z,u)}
W^{\phi_1\phi_2}_{(u,x_2)}\, ,
\ee
which is valid for any $\Omega$, however thorough this paper 
$\Omega$ always stands for  some $R$. Note that eq. (\ref{srpen}) simply
follows from the identity (\ref{iden}).
Taking $J,R=0$ then  the obtained integrals in eqs. (\ref{jednicka}),(\ref{dvojka})
correspond with the third and the fourth diagram in the diagrammatic expression 
for selfenergy (\ref{selfa}).

The terms which follow  from the quartic interactions are
\bea    \label{trojka}
&&\frac{\dl \check{A}^3(x)}{\dl R_A(y)}1=
\frac{\dl }{\dl R_A(y)}
\left[R_A^3(x)-3i R_A(x) W^{AA}_{(x,x)}\right.
 \\
&&\left.+(-i)^2\int \tilde{dt}\tilde{dz}\tilde{du} W^{AA}_{(x,t)}W^{AA}_{(x,z)} 
\Gamma^{R_AR_AR_A}_{(t,z,u)}W^{AA}_{(u,x)}\right]
\nn \\
&&=3R_A^2(x)\dl_{xy}-3i W^{AA}_{(x,x)}\dl_{xy}
\nn \\
&&-3i R_A(x)  \tilde{dz}\tilde{du} W^{AA}_{(x,z)} \Gamma^{R_AR_AR_A}_{(y,z,u)}W^{AA}_{(u,x)}
\nn \\
&&+(-i)^2\int \tilde{dt}\tilde{dz}\tilde{du} W^{AA}_{(x,t)}W^{AA}_{(x,z)} \Gamma^{R_AR_AR_AR_A}_{(y,t,z,u)}W^{AA}_{(u,x)}
\nn \\
&&+3(-i)^2\int  \tilde{dr}\tilde{ds}\tilde{dt}\tilde{dz}\tilde{du}
W^{AA}_{(x,r)}\Gamma^{R_AR_AR_A}_{(y,r,s)}W^{AA}_{(s,t)}
W^{AA}_{(x,z)}\Gamma^{R_AR_AR_A}_{(t,z,u)}W^{AA}_{(u,x)};
\nn
\eea
%
\bea  \label{ctyrka}
&&\frac{\dl \check{A}(x)\check{B}^2(x)}{\dl R_A(y)}1=
\frac{\dl }{\dl R_A(y)}
\left[R_A(x)R_B^2(x)-i R_A(x) W^{BB}_{(x,x)}\right.
 \\
&&\left.+(-i)^2\int \tilde{dt}\tilde{dz}\tilde{du} W^{AA}_{(x,t)}W^{BB}_{(x,z)} 
\Gamma^{R_AR_BR_B}_{(t,z,u)}W^{BB}_{(u,x)}\right]
\nn \\
&&=R_B^2(x)\dl_{xy}-i W^{BB}_{(x,x)}\dl_{xy}
\nn \\
&&-i R_B(x) \int \tilde{dz}\tilde{du} W^{AA}_{(x,z)} \Gamma^{R_AR_BR_B}_{(y,z,u)}W^{BB}_{(u,x)}
\nn \\
&&+(-i)^2\int \tilde{dt}\tilde{dz}\tilde{du} W^{AA}_{(x,t)}W^{BB}_{(x,z)} \Gamma^{R_AR_AR_BR_B}_{(y,t,z,u)}W^{BB}_{(u,x)}
\nn \\
&&+2(-i)^2\int \tilde{dr}\tilde{ds}\tilde{dt}\tilde{dz}\tilde{du} W^{BB}_{(x,s)}\Gamma^{R_AR_BR_B}_{(y,s,t)}W^{BB}_{(t,z)}
 W^{AA}_{(x,r)}\Gamma^{R_AR_BR_B}_{(r,z,u)}W^{BB}_{(u,x)}
\nn\\
&&+(-i)^2\int \tilde{dr}\tilde{ds}\tilde{dt}\tilde{dz}\tilde{du}
 W^{AA}_{(x,s)}\Gamma^{R_AR_AR_A}_{(y,s,t)}W^{AA}_{(t,r)}
 W^{BB}_{(x,z)}\Gamma^{R_AR_BR_B}_{(r,z,u)}W^{BB}_{(u,x)}/, .
\nn
\eea

Setting  $R$ to zero  then the first line after the second equality of the above 
expressions (\ref{trojka}),(\ref{ctyrka}) is responsible for generation of the 
tadpole diagrams wherein the vertex is undressed. The third lines of the above 
expressions correspond
with the sunset diagrams wherein one quartic vertex is dressed. Considering the 
prefactors  the sunset diagrams contribution to $\Pi(x,y)$
is as following
\bea \label{sunset}
&&-(i)^2 2g^2\int \tilde{dl}\tilde{dq}\tilde{dp} e^{ip.(x-y)} G_A(p-l)G_A(l-q)G_A(q)\Gamma_{A^4}(p;q,l-q,p-l)
\nn \\
&&-(i)^2 2g^2\int  \tilde{dl}\tilde{dq}\tilde{dp} e^{ip.(x-y)} G_B(p-l)G_B(l-q)G_A(q)\Gamma_{A^2B^2}(p;q,l-q,p-l)
\nn
\eea
where $\Gamma_{A^4},\Gamma_{A^2B^2}$ are the exact proper quartic vertices. 
The momentum corresponding to  external leg of the full vertex is ingoing, while all the momenta associated with lines inside the loop are outgoing from this vertex. 

The three remaining contribution in (\ref{trojka}),(\ref{ctyrka}) corresponds with twoloop diagram of the same topology. The last term in (\ref{trojka}) corresponds with the last diagram in 
(\ref{selfa}), while the both remaining terms in (\ref{ctyrka}) correspond to one to last diagram in rel. (\ref{selfa}).  

The last term in eq. (\ref{xadse2}) corresponds with the fermion loop diagram depicted  in (\ref{USDE}). Suppressing dirac indices the appropriate contribution to $\Pi_A$ reads
\be
\frac{\delta \bar{\psi}(x)\psi(x)}{\delta A(y)}= i\, Tr\,  \int \tilde{dz}\tilde{du} 
\Gamma^{\eta\eta R_A}_{(y,z,u)}W^{\psi\psi}_{(x,z)}W^{\psi\psi}_{(u,x)}\, .
\ee

The derivation of higher vertices proceed through the further differentiation with 
respect of a given $R$'s. However in this paper we need to know only  classical 
(tree-level) vertices, zero order of the  Planck constant expansion result must
 correspond with the variation of the classical action with respect to the a 
given original fields, i.e. $A,B,\psi$. Using the obvious and convenient notation  
 $\left.\Gamma_{A^kB^n..}(x_1...x_{n+k}..)=
\Gamma^{R_{A_1}...R_{B_{n}}..}_{(x_1...x_{n+k}..)}\right|_{R=0}$
 we list the appropriate vertices here  for completeness:     
%
%
%
\bea \label{stromy}
\Gamma_{A^3}(x,y,z)=-6mg\dl_{xy}\dl_{yz};&&
\quad \Gamma_{AB^2}(x,y,z)=-2mg\dl_{xy}\dl_{yz}
\nn\\
\Gamma_{\psi\psi A}(x,y,z)=-2g\dl_{xy}\dl_{yz};&&
\Gamma_{\psi\psi B}(x,y,z)=-2g\dl_{xy}\dl_{yz};
\nn \\
\Gamma_{A^4}(x,y,z,t)=-12g^2\dl_{xy}\dl_{yz}\dl_{zt};&&
\Gamma_{A^2B^2}(x,y,z,t)=-4g^2\dl_{xy}\dl_{yz}\dl_{zt};
\nn \\
\Gamma_{B^4}(x,y,z,t)=-12g^2\dl_{xy}\dl_{yz}\dl_{zt}&&\, \, .
\eea
Substituting the vertices (\ref{stromy}) into their right places, 
we can write down the resulting expression for the function $\Pi_A$.
For practical purposes we Fourier transform the result into the momentum space.
After the little algebra we get
\bea  \label{analog}
&&\Pi_A(p)=-ig^2\int \tilde{dl}
\biggl\{-8\frac{{\cal{F}}(l){\cal{F}}(q)}{l^2-{\cal{M}}^2(l)}
-4\frac{{\cal{F}}(l){\cal{F}}(q)[-p^2+({\cal{M}}(l)+{\cal{M}}(q))^2]}
{(l^2-{\cal{M}}^2(l))(q^2-{\cal{M}}^2(q))}
\nn \\
&&+6 G_A(l)+2 G_B(l)
+18m^2 G_A(l)G_A(l-p)
+2m^2 G_B(l)G_B(l-p)\biggr\}\, ,
\eea   
where the second line follows from bosonic loops while the first line follows from the 
fermionic loop and where we used notation $q=p-l$.

The remaining SDEs can be derived from (\ref{qoe}) by very similar fashion as the equation 
for 
$G_A$. From the following definition 
\be
G_B^{-1}(x,y)=\left.\frac{\dl^2\Gamma[R]}{\dl R_B(y)\dl R_B(x)}\right|_{R=0}
\ee
we can obtain the equation for the selfenergy function $\Pi_B$.
In our approximation of SDEs the result reads
\bea \label{katalog}
G_B^{-1}(p)&=&p^2-m^2-\Pi_B(p)
\nn \\
\Pi_B(p)&=&-ig^2\int \tilde{dl} \biggl\{6 G_B(l)+2 G_A(l)
+4m^2 G_A(l)G_B(l-p)
\nn \\
&&-8\frac{{\cal{F}}(l){\cal{F}}(q)}{l^2-{\cal{M}}^2(l)}
-4\frac{{\cal{F}}(l){\cal{F}}(q)[-p^2+({\cal{M}}(l)-{\cal{M}}(q))^2]}
{(l^2-{\cal{M}}^2(l))(q^2-{\cal{M}}^2(q))}\biggr\}
\eea   

The majorana fermion mass function ${\cal{M}}$ as well as the renormalization wave-function 
${\cal{F}}$ in  the bare vertex approximation of  the fermion SDE are 
\bea  \label{dialog}
\frac{1}{{\cal{F}}(p)}&=&1+i4g^2\int\tilde{dl}\biggl\{ \frac{{\cal{F}}(l)\frac{p.l}{p^2}}{l^2-{\cal{M}}(l)}
\left[G_A(l)+G_B(q)\right]\biggr\}
\nn \\
\frac{{\cal{M}}(p)}{{\cal{F}}(p)}&=&m+i4g^2\int\tilde{dl}\biggl\{ \frac{{\cal{F}}(l){\cal{M}}(l)}{l^2-{\cal{M}}^2(l)}
\left[G_A(q)-G_B(q)\right]\biggr\}\, ,
\eea
noting that there are only one loop diagrams even in the exact case.  


\section{Numerical  results for SDEs of WZM}

In this section we present numerical results for the propagators
as they have been obtained from the solution of the SDEs.
To obtain this, we convert SDEs  from Minkowski space
into the Euclidean space via performing Wick rotation.
The resulting set of rotated equations is presented in the Appendix A for a
reader's convenience. Here, we should mention about
possibility to solve SDEs directly in Minkowski space \cite{SAULI2005}, 
however this far nontrivial
 approach is until now  developed for renormalized theory only.
Therefore at this stage of our calculation we deal with Euclidean formalism, 
which method  leaves us with the solutions for space-like momenta only.
                                                                                                                             
Before describing a quantitative face of the model we anticipate basic results. 
The most striking fact is the unexpected behavior of the boson 
selfenergy functions $\Pi_{A,B}$. In accordance
of our expectation they are degenerated,
i.e. the equation  $\Pi_A=\Pi_B$ is valid (with  negligibly small
numerical error). On the other the boson selfenergy appears
to be fairly unrelated with the fermion one. 
The already perturbative
observed violation from Susy Ward identity is further enhanced.
Again we will present the solution for unrenormalized but regularized Green functions.
(Notation: solving the renormalized set of SDEs one can always  approximately preserve
Ward identity at least in the vicinity of the renormalization. 
This is not case we are interested in.) 
                                                                                                                             
Susy invariant regularization of {\bf I} was already proposed in the paper  
\cite{ILIZUM1974}. The regulator Lagrangian term of the form
\be
\frac{{\cal L}_{kin}(\phi \rightarrow \partial_{\mu}\partial^{\mu}\phi)}{\Lambda^2}
\ee
has been added into the original Lagrangian {\bf I} ( ${\cal L}_{kin}$
corresponds with the first line
of \ref{clasI}). This approach -albeit useful in perturbation theory analyzes
in  the limit $\Lambda\rightarrow\infty$-
becomes extremely inconvenient in nonperturbative study like here, especially
when $\Lambda$ remains finite. In fact, due to the number of  time differentiation,
new degrees of freedom appear and it is rather nontrivial task
to recognize particle content of such theory
( associated with the  S-matrix
singularity rather then the number of the fields in the Lagrangian).
However, the theory {\bf II}  does not  respect Susy
due to its radiative corrections breaking. We are not so restricted by the
choice of the regularization scheme and we use two
very simple regulator functions in our set of  Euclidean SDEs. As a first we have explored
hard cutoff, i.e. the momentum integrals in \ref{dysoni} become
\be
\int dx \rightarrow \int dx \theta(\Lambda^2-x)\, .
\ee
As a second prescription we use the smooth regulator function, such that
\be
\int dx \rightarrow \int dx \frac{1}{(x-\Lambda^2)(z-\Lambda^2)}
\label{regulat}
\ee
($x,z$ are the squares of Euclidean momenta defined below eqs. (\ref{dysoni})).
                                                                                                                             
Up to a few percentages accuracy,  we have found, these two regularization schemes become
identical for $\Lambda>>m$. The presented numerical results were calculated within the
use of  second approach, i.e. with regulator (\ref{regulat}).

\begin{figure}
\label{mesheffect}
\centerline{\mbox{\epsfig{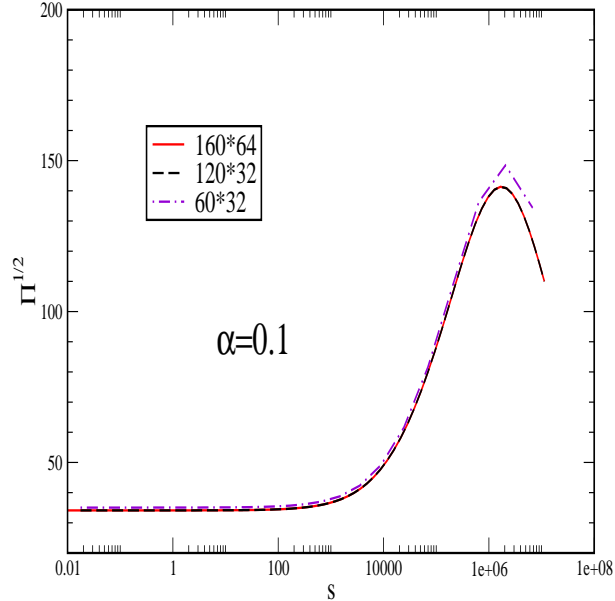}
}} \caption[caption]{Test of numerical precision. 
Square root of the boson selfenergy $\Pi_{A,B}(p^2)$ 
as has been obtained with various number of mesh-points, i.e.
({\it number  of discretized momentum points*number of discretized point of $\cos\theta$}) 
Here $\Lambda^2=10^6m^2$. }
\end{figure}

\begin{figure}
\label{fmajorana}
\centerline{\mbox{\epsfig{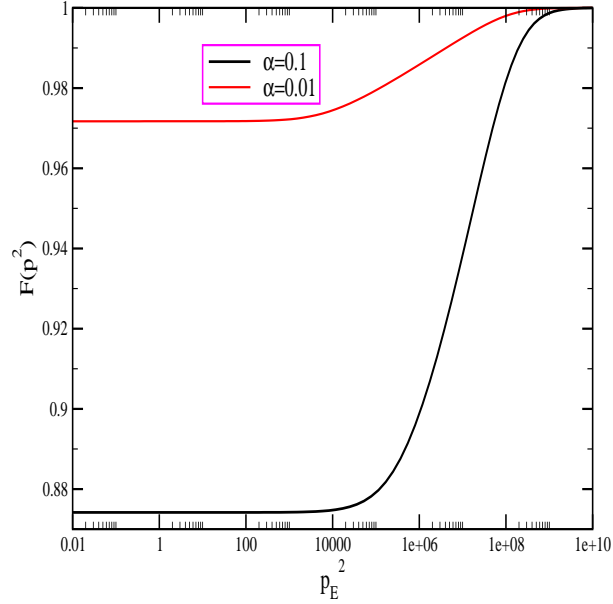}
}} \caption[caption]{Renormalization wave function ${\cal F}$ for two values of the coupling constant $\alpha=\frac{g^2}{4\pi}$ and ultraviolet cutoff $\Lambda^2=10^8m^2$. $m$ is the Lagrangian mass. }
\end{figure}

\begin{figure}
\label{cutoff}
\centerline{\mbox{\epsfig{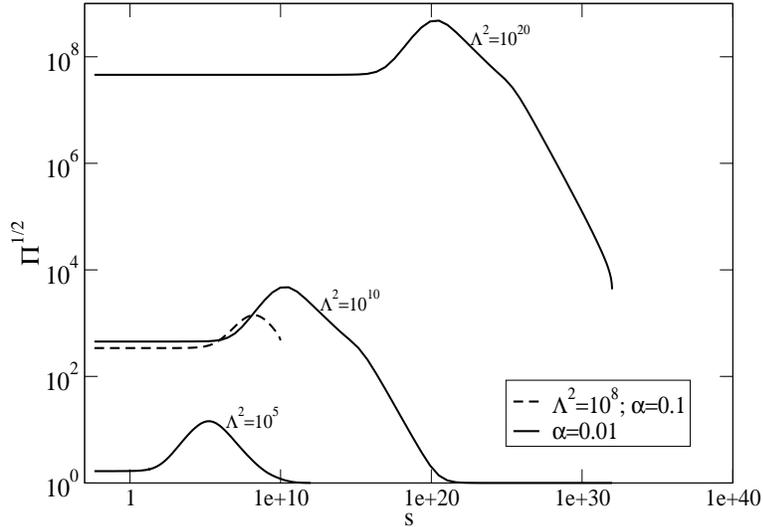}
}} \caption[caption]{Square root of the boson selfenergy $\Pi_{A,B}(p^2)$ for $\alpha=0.01$. Figure shows the dependence on the ultraviolet cutoff $\Lambda$. The result for $\alpha=0.1$ and 
$\Lambda^2=10^8m^2$ is added or comparison.  The results have physical meaning only bellow the cutoff ,i.e. on the left side from the maximum of  a given line.}
\end{figure}

To integrate the equations  numerically  we use Gaussian quadrature method. 
In the case of usage of regulator function (\ref{regulat}) the largest momentum was 
typically $p>>\Lambda$ (while the obtained functions have physical 
only in the range $p<\Lambda$).   
The equations (\ref{dysoni}) have been solved by the method of iterations.
In this treatment we need a knowledge about the functions on a grid which differ from
the  one we choose to discretize of momenta $y\rightarrow y_i$. For this purpose we
interpolate the appropriate functions inside the kernels of SDEs. The numerically 
convergence was observed already for relatively small number of mesh points. 
This is shown in the figure \ref{mesheffect}. The renormalization wave function
${\cal F}$ is displayed in the figure \ref{fmajorana}, noting that due to the degeneracy of scalar and pseudoscalar the mass function is fully given by its inverse 
i.e.${\cal M}=m/{\cal F} $. The boson selfenergy is shown in the figure (\ref{cutoff}).
For a very large cutoff it  behaves like a constant in the low energy where 
$s=p^2<<\Lambda^2$. The results were calculated for rather small coupling $\alpha=0.01$
and for three different values of cutoff $\Lambda^2/m^2=10^5,10^{10},10^{20}$.


\section{Conclusion and speculations}
We studied quantum corrections to the propagators of Susy model without the explicit presence 
of auxiliary fields. In the section 3 we have shown that the next to leading corrections correction  lead to different prediction then known from  manifest 
Susy invariant treatment theory {\bf I}. 
Calculated Green's functions  do not respect Susy Ward identity 
own to the formalism {\bf I}. The origin of this can be easily traced: the  
preferable choice of unperturbed Lagrangian \ref{clasI}
does not correspond with with free Lagrangian {\bf II}. 

In addition we have shown that this 'unexpected discrepancy' between {\bf I} and {\bf II}
(however it was somehow indirectly  intimate already 
in the paper \cite{WESZUM1974}) becomes  even more crucial  when the solution of  SDEs is considered. Of course, the appropriate solution is only certain approximation of the  full nonperturbative solution (e.g. it is truncation of SDEs system dependent), however, in agreement with  experience from another quantum field models, there is no large chance for some significant changes when further  improvement will be made. Some lattice simulation, recently undone, but hopefully made in the near future could outshine shades at this area (for some progress see the paper \cite{FEO}). The observed violation of non-renormalization theorem
has further non-trivial consequences. This is because the coupling constants of cubic interactions
(in {\bf II}) are proportional to bare Lagrangian mas $m$. These need to be specified when dealing with renormalization. The already mentioned lattice formulation of the problem could be actually credible guide. At this place we should mention the paper \cite{BARKRA1983}  where the authors observed the splitting of the renormalized coupling constant as 
$g\rightarrow g_{AAF},g_{AGB^2},g_{A\psi\psi}$ (remind, the authors of this paper dealt with {\bf I}). This is an another indication that a naive formal quantization and associated functional method applied on auxiliary fields (i.e. the field without canonical conjugate momenta) can fail in prediction when compared to the result obtained from the sophisticated  first principle method (here directly from the discretized path integral).

Although the model studied here has obviously no phenomenological interest, it has a large historical impact on a part of particle physics. At given time an extreme effort and many speculations have been  pursued to uncover and explain  possible mechanism of Susy breaking 
\cite{FAYILI1975,ZUMINO1975,FAYET1975,WITTEN,GIRGRI1982},... . It seems very likely that such troubles can be overcome if  one use honestly constrain ,i.e. equation of motions for auxiliary field content, before the quantization and thus then deal with the usual quantum field theory expanded around the classical vacuum,  where the nearest excitation are  just free physical particles.   

{\bf Acknowledgments}

The work was partially  supported by the grant GA CR 202/03/0210 and by the ASCR K1048102.

\newpage
\section{Appendix A}

\begin{center}{\bf SDEs for WZM in Euclidean space}\end{center}

Converting (\ref{analog},\ref{katalog},\ref{dialog})
to the Euclidean space ($l_0=il_{4E}$) we should obtain

\bea  \label{dysoni}
\Pi_A(y)&=&\frac{\alpha}{2\pi^2}\int_0^{\infty} dx 
\int_0^{\pi}d\theta x\sin^2\theta{\cal{J_A}}\quad ; \quad
\Pi_B(y)=\frac{\alpha}{2\pi^2}\int_0^{\infty}  dx
\int_0^{\pi}d\theta x\sin^2\theta{\cal{J_B}}
\nn \\
{\cal{J_A}}&=&\biggl\{6 G_A(x)+2 G_B(x)
-18m^2 G_A(x)G_A(z)
-2m^2 G_B(x)G_B(z)
\nn \\
&&-8\frac{{\cal{F}}(x){\cal{F}}(z)}{x+{\cal{M}}^2(x)}
+4\frac{{\cal{F}}(x){\cal{F}}(z)[y+({\cal{M}}(x)+{\cal{M}}(z))^2]}
{(x+{\cal{M}}^2(x))(z+{\cal{M}}^2(z))}\biggr\}
\eea

\bea
{\cal{J_B}}&=&\biggl\{6 G_B(x)+2 G_A(x)
-4m^2 G_A(x)G_B(z)
\nn \\
&&-8\frac{{\cal{F}}(x){\cal{F}}(z)}{x+{\cal{M}}^2(x)}
+4\frac{{\cal{F}}(x){\cal{F}}(z)[y+({\cal{M}}(x)-{\cal{M}}(z))^2]}
{(x+{\cal{M}}^2(x))(z+{\cal{M}}^2(z))}\biggr\}
\eea

\bea
\frac{1}{{\cal{F}}(y)}&=&1+\frac{2\alpha}{\pi^2}\int_0^{\infty} dx 
\int_0^{\pi}d\theta x\sin^2\theta
\biggl\{ \frac{{\cal{F}}(l)\sqrt{(x/y)}\cos\theta}{x+{\cal{M}}^2(x)}
\left[G_A(z)+G_B(z)\right]\biggr\}
\nn \\
\frac{{\cal{M}}(y)}{{\cal{F}}(y)}&=&m+\frac{2\alpha}{\pi^2}\int_0^{\infty} dx 
\int_0^{\pi}d\theta x\sin^2\theta
\biggl\{ \frac{{\cal{F}}(x){\cal{M}}(x)}{x+{\cal{M}}^2(x)}
\left[G_A(z)-G_B(z)\right]\biggr\}
\eea

where $z=y-2\sqrt{xy}\cos\theta+x$,
$G_{A,B}(x)=[x+m^2+\Pi_{A,B}(x)]^{-1}$


\begin{thebibliography}{99}
\bibitem{notation}
we use Dirac matrices $\{\gamma^{\mu},\gamma^{\nu}\}=2g^{\mu\nu}$ and Minkowski metric
$g_{\mu\nu}=diag(1,-1,-1,-1)$
\end{thebibliography}
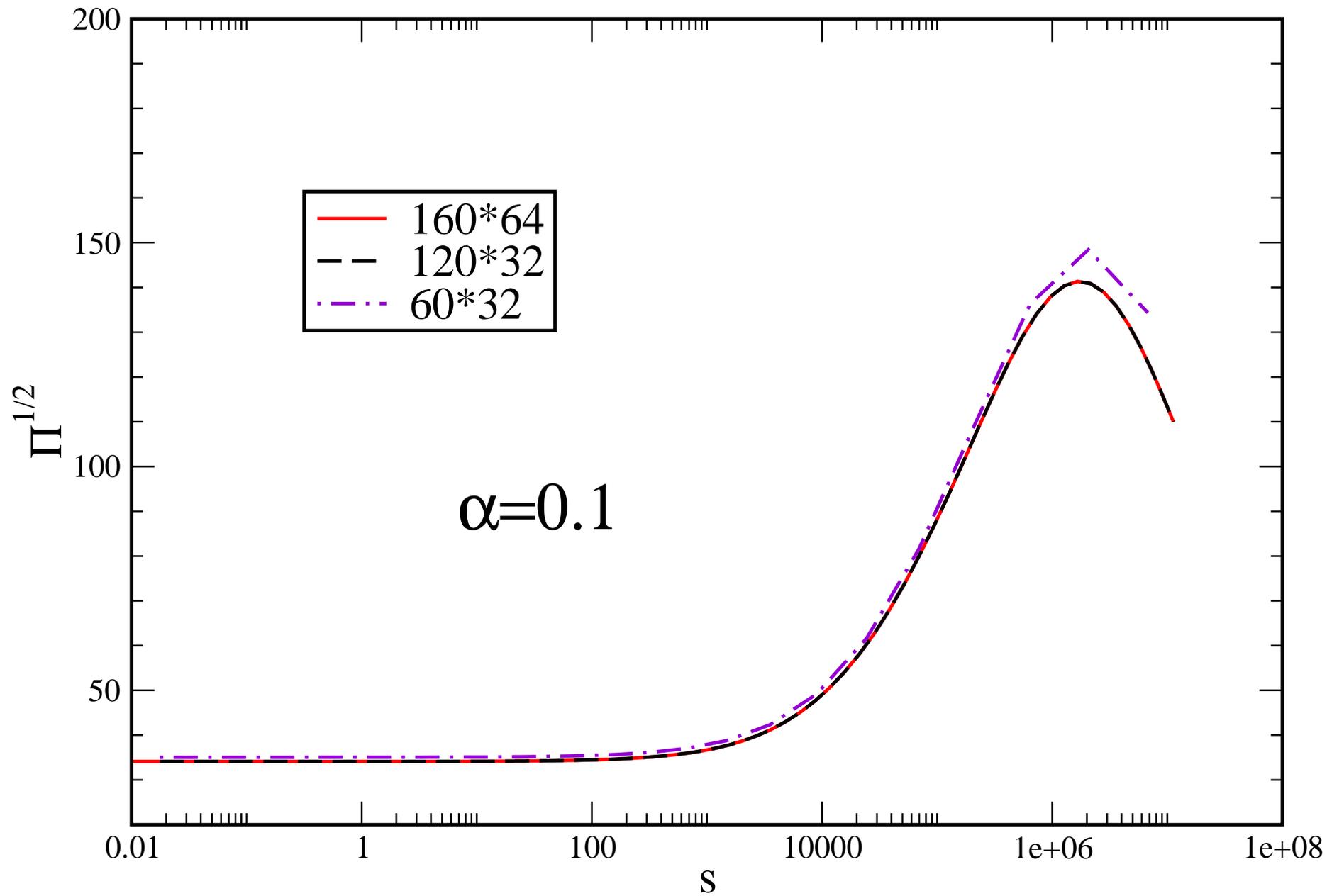
\end{document}